\documentclass{iau}
\usepackage{graphicx,natbib,amsmath}

 \def\vss{\vskip 6pt} 
\def\makeheadline{\vbox to 0pt{\vskip-30pt\line{\vbox to8.5pt{}\the
                               \headline}\vss}\nointerlineskip}

\def\footnoterule{\kern-3pt \hrule width \hsize \kern 2.6pt \vskip 3pt}

\def\ts{\thinspace}

       \def\bk{\kern -0.3em}  \def\b{\kern -0.1em}
\def\etal{et~al.\ }  
\def\gapprox{$_>\atop{^\sim}$}  \def\lapprox{$_<\atop{^\sim}$}

\newdimen\sa  \def\sd{\sa=.1em  \ifmmode $\rlap{.}$''$\kern -\sa$
                                \else \rlap{.}$''$\kern -\sa\fi}
              \def\dgd{\sa=.1em \ifmmode $\rlap{.}$^\circ$\kern -\sa$
                                \else \rlap{.}$^\circ$\kern -\sa\fi}
\newdimen\sb  \def\md{\sa=.06em \ifmmode $\rlap{.}$'$\kern -\sa$
                                \else \rlap{.}$'$\kern -\sa\fi}
\title[Scaling Laws for Dark Matter Halos]{Scaling Laws for Dark Matter Halos in Late-Type and Dwarf Spheroidal Galaxies}

\author[Kormendy \& Freeman]{John Kormendy$^1$ \and K. C. Freeman$^2$}

\affiliation{$^1$Department of Astronomy, University of Texas at Austin,\\
        2515 Speedway, Mail Stop C1400,
        Austin, Texas 78712-1205, USA,\\
        email: {\tt kormendy@astro.as.utexas.edu;}\\
        Max-Planck-Institut f\"ur Extraterrestrische Physik,\\
        Giessenbachstrasse, D-85748 Garching-bei-M\"unchen, Germany;\\
        Universit\"ats-Sternwarte, Scheinerstrasse 1, D-81679 M\"unchen, Germany;\\
        ~~~~~~~~~~~~\\
        $^2$Research School of Astronomy and Astrophysics,\\
        Mount Stromlo Observatory, The Australian National University,\\ 
        Cotter Road, Weston~Creek, Canberra, ACT 2611, Australia,\\
        email: {\tt Kenneth.Freeman@anu.edu.au}}

\pubyear{2014}
\volume{311}
\jname{Galaxy Masses as Constraints of Formation Models}
\editors{M. Cappellari \& S. Courteau, eds.}
\begin{document}

\maketitle

\begin{abstract}
Dark matter (DM) halos of Sc{\ts}--{Im} galaxies satisfy structural scaling laws analogous 
to the fundamental plane relations for elliptical galaxies.  Halos in less luminous galaxies 
have smaller core radii $r_c$, higher central densities $\rho_\circ$, and smaller central velocity 
dispersions $\sigma$.  If dwarf spheroidal (dSph) and dwarf Magellanic irregular (dIm) galaxies
lie on the extrapolations of these correlations, then we can estimate their baryon loss  relative
to that of Sc{\ts}--{\ts}Im galaxies.  We find that, if there had been no enhanced baryon loss
relative to Sc{\ts}--{\ts}Im galaxies, typical dSph and dIm galaxies would be brighter by 
$\Delta M_B \simeq -4.0$ mag and $\Delta M_B \simeq -3.5$ mag, respectively.  Instead, the galaxies
lost or retained as gas (in dIm galaxies) baryons that could have formed stars.  Also, dSph and dIm
galaxies have DM halos that are more massive than we thought, with $\sigma \sim 30$ km s$^{-1}$ or 
circular-orbit rotation velocities $V_{\rm circ} \sim 42$ km s$^{-1}$.  Comparison of DM and visible
matter parameter correlations confirms that, at $M_V$\ts\gapprox\ts$-18$, dSph and dIm galaxies form 
a sequence of decreasing baryon-to-DM mass ratios in smaller dwarfs.  We show explicitly that galaxy 
baryon content goes to (almost) zero at $V_{\rm circ}$\ts\lapprox\ts42\ts\ts$\pm$\ts4{\ts\ts}km{\ts\ts}s$^{-1}$, 
in agreement with~$V_{\rm circ}$ as found from our estimate of baryon depletion.  Our results suggest
that there may be a large population of DM halos that are dark and undiscovered.  This helps to solve
the problem that the initial fluctuation spectrum of cold dark matter predicts more dwarf galaxies
than we observe.
\end{abstract}

\firstsection
\section{Introduction and Analysis Machinery}

      This paper summarizes Kormendy \& Freeman (2014).  That paper derives structural parameter 
correlations for DM halos in Sc--Im and dSph galaxies.  We restrict ourselves to objects that contain 
only two main components, a baryonic disk or main body and a DM halo.  For galaxies with well measured 
H{\ts}I rotation curves $V(r)$, the derived DM parameters come from published maximum-disk decompositions 
of $V(r)$ into visible and dark components.  The halo model used is the nonsingular isothermal.
At absolute magnitude $M_B$\gapprox$-14$, $V$ is comparable to the velocity dispersion; then rotation 
curve decomposition is impossible.  For these dwarf spheroidal (dSph) and dwarf Magellanic irregular 
(dIm) galaxies, we derive central DM densities using the Jeans equation.

      We compare DM parameters with visible matter parameter correlations from Kormendy
\& Bender (2012).  They show that the parameter correlations of Sph galaxies are continuous
with the disks (but not bulges) of S0 galaxies.  In essence, Sph galaxies are bulgeless S0s.  Moreover,
Sph and S$+$Im galaxies have similar structure at each $M_V$.  Effective brightnesses decrease
dramatically at $M_V$\ts$>$\ts$-18$; this is used in Figure\ts3 here.

\section{DM Structural Parameter Correlations}

      Figures 1 and 2 show the main observational result of Kormendy \& Freeman (2014):
DM halos of Sc\ts--{\ts}Im galaxies satisfy well defined scaling~laws.  Halos in less luminous galaxies 
have smaller core radii, higher central densities, and smaller velocity dispersions.  This confirms 
previous analyses of smaller samples (Kormendy 1988, 1990; Kormendy \& Freeman 2004).  Scaling laws provide 
new constraints on galaxy formation.  For example:

\vfill

\begin{figure}[h]

\includegraphics{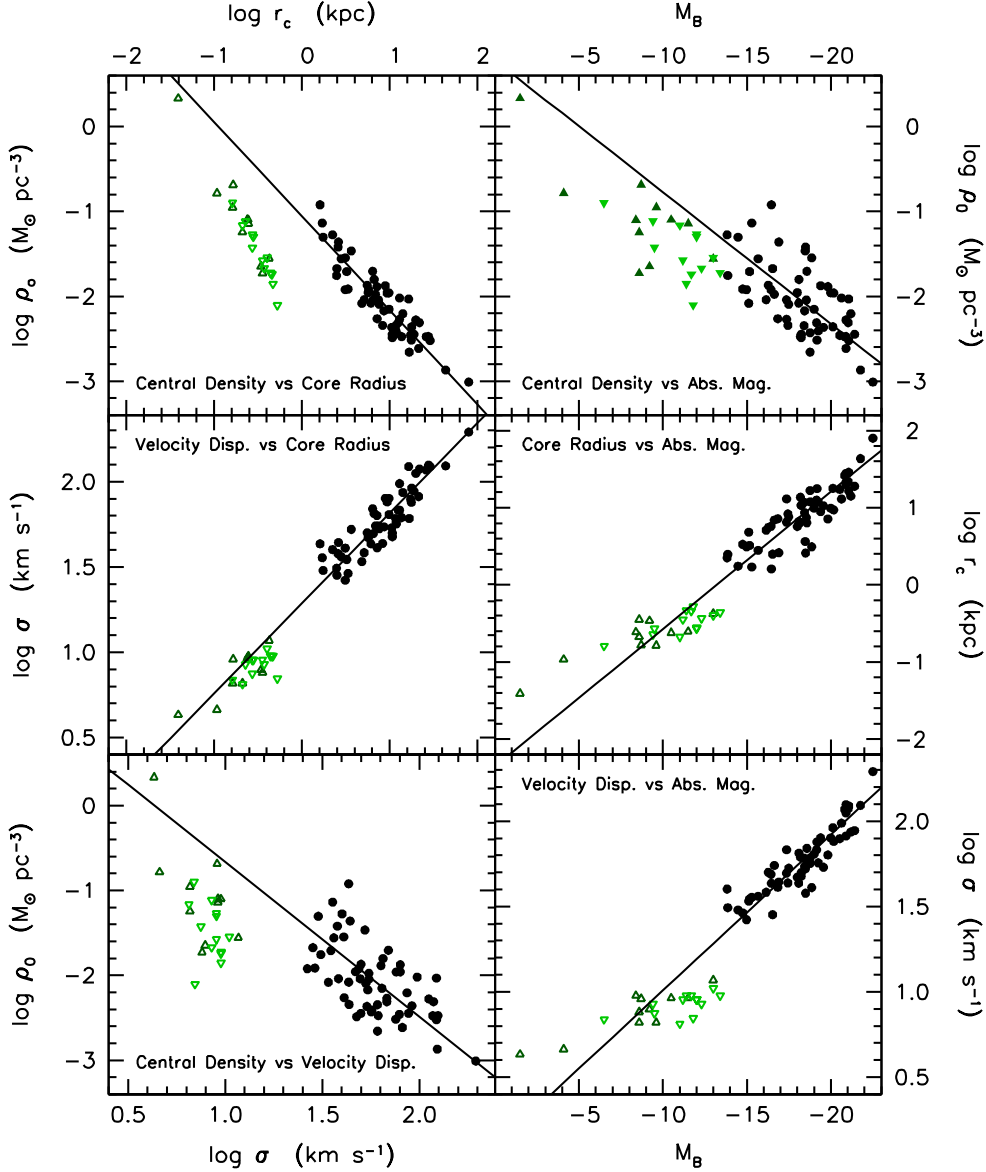}

\caption{\lineskip=0pt \lineskiplimit=0pt
Dark matter parameter correlations for Sc{\thinspace}--{\thinspace}Im galaxies as derived from rotation curve
decompositions using the nonsingular isothermal as DM model ({\it black points and black line = symmetric
least-squares fit\/}).  Also added are dSph galaxies ({\it green triangles}) and dIm galaxies ({\it upside-down 
green triangles}).  For dSph and dIm galaxies, $\rho_\circ$ is a meaure of the DM and therefore is plotted 
with filled symbols in the top-right panel.  But $r_c$ and $\sigma$ are visible-matter parameters and so are 
plotted with open symbols in the other panels.  From Kormendy \& Freeman (2014).
}

\end{figure}

\eject

     Halo density depends on collapse redshift $z_{\rm coll}$ as \hbox{$\rho_\circ \propto (1 + z_{\rm coll})^3$}.
Thus~$\rho_\circ$ increases toward lower luminosities because fainter galaxies collapsed earlier.  
The DM correlations imply that dwarf galaxies formed at least $\Delta\,z_{\rm coll} \simeq 7$ 
earlier than giant spirals.  Correction for baryonic DM compression would make the ``pristine'' 
$\rho_\circ$ smaller for giant galaxies and would increase $\Delta\,z_{\rm coll}$.

\vfill

\begin{figure}[h]

\includegraphics{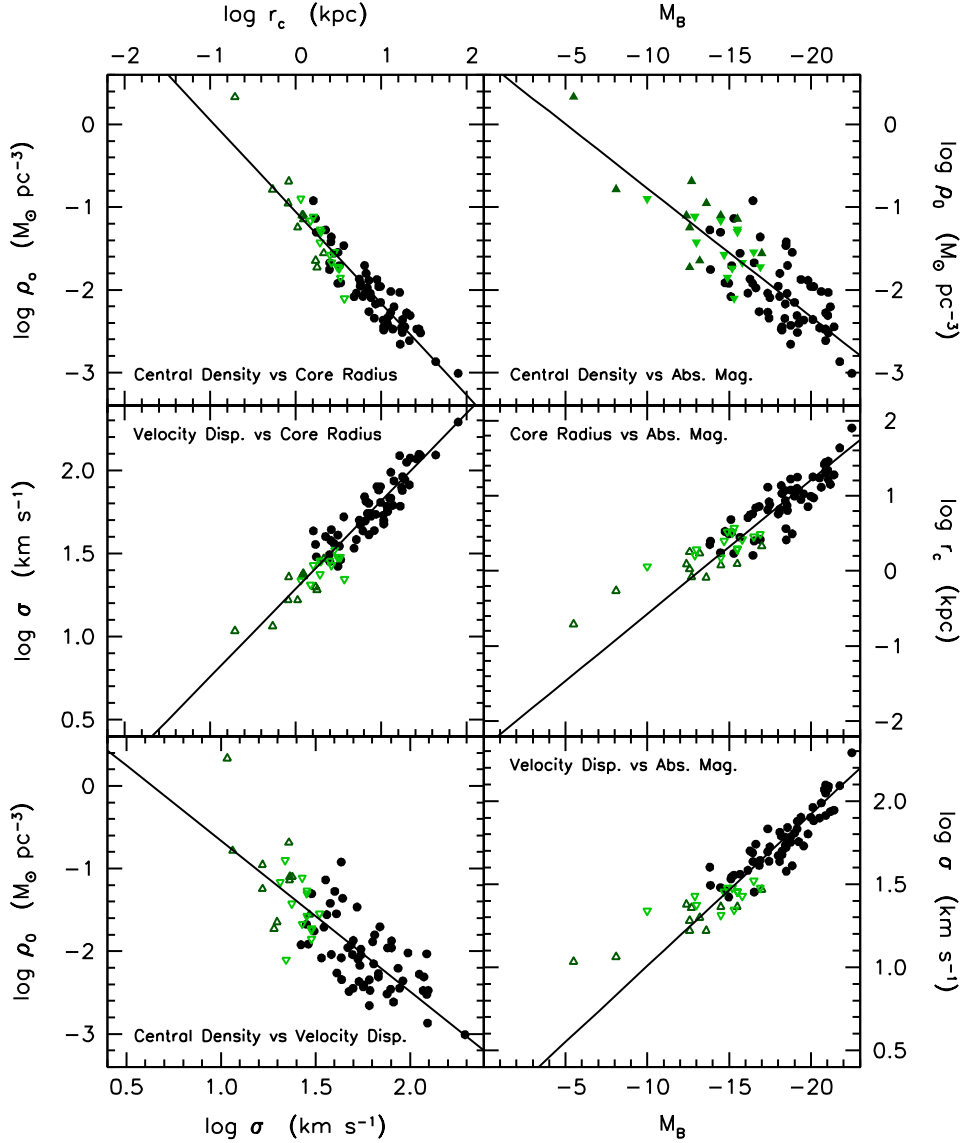}

\caption{\lineskip=0pt \lineskiplimit=0pt
DM correlations for  Sc{\thinspace}--{\thinspace}Im, dSph, and dIm galaxies (Figure 1)
but~with~dSph and dIm galaxies shifted in $M_B$, $\log r_c$, and $\log \sigma$ (but not $\log \rho_\circ$)
to move them onto the scaling laws for the Sc{\thinspace}--{\thinspace}Im galaxies.~The goal is to estimate 
(1) the baryon~loss from dwarf galaxies relative to that from Sc{\ts}--{\ts}Im galaxies and 
(2) the difference between the baryon and DM values of 
$\log r_c$ and $\log \sigma$. The shifts for dSph and dIm galaxies are, respectively,  
$M_B \rightarrow M_B - 4.0$ and $M_B \rightarrow M_B - 3.5$, 
$\log r_c \rightarrow \log r_c + 0.70$ and $\log r_c \rightarrow \log r_c + 0.85$, and 
$\log \sigma \rightarrow \log \sigma + 0.40$ and $\log \sigma \rightarrow \log \sigma + 0.50$.
Note the interpretation of the shifts: $\Delta \log{r_c}$ gives us the ratio of DM core radius to 
visible matter core radius, and $\Delta \log{\sigma}$ gives us the ratio of DM velocity dispersion 
to visible matter velocity dispersion.  From Kormendy\ts\&{\ts}Freeman\ts(2014).
}

\end{figure}

\eject

      In Figure\ts3, smaller dwarfs have smaller stellar-to-DM ratios; these are probably 
associated with baryon loss.  The correlations in Figure\ts1 provide a way to estimate this 
baryon loss and the properties of dwarf galaxy halos. The Jeans equation tells us the DM central 
density $\rho_\circ$ but not its core radius $r_c$ or velocity dispersion $\sigma$. 
So the top-right panel in Figure\ts1 correctly shows an offset 
of dSph+dIm galaxies from the extrapolation of the fitted \hbox{$\rho_\circ$--$M_B$} correlation. 
But the smallest galaxies with rotation curve decompositions have DM densities similar to those of
the biggest dwarfs with Jeans equation estimates. {\it We assume that dSph+dIm galaxies 
would lie on the extrapolation of the $\log \rho_\circ - M_B$ correlation for bright galaxies 
except for the effects of their enhanced baryon loss.}  The top-right panel of Figure\ts1 then tells 
us the $\Delta M_B$ shifts that bring dSph and dIm galaxies onto the fitted relation.  The top-left 
and bottom-left panels tell us the shifts in $\log r_c$ and $\log \sigma$ that bring dSph$+$dIm 
galaxies onto those relations. These shifts are applied in Figure\ts2.  In all panels, dSph and dIm 
galaxy halos lie on the correlations for more massive galaxies.  The $\Delta \log{\sigma}$ shifts imply that
almost-dark dwarfs are more massive~than we thought.  Their typical halo has $\sigma \sim 30$ 
km s$^{-1}$.  This corresponds to $V_{\rm circ} \sim 42$ km s$^{-1}$, 
in remarkably good agreement with the value of $V_{\rm circ}$ where galaxies get dim in Figure\ts4.

\section{Comparison of Scaling Relations for Visible and Dark Matter}

      Figure 3 compares the central projected densities of DM halos with effective projected densities 
of stars.  S$+$Im$+$Sph galaxies with $M_V$ \gapprox \ts$-18$ form a sequence of decreasing 
baryon-to-DM density ratios at decreasing $L_V$.  We suggest that they form a sequence of decreasing 
baryon retention (vs.~supernova-driven winds:~Dekel \& Silk 1986) or decreasing baryon capture (after 
cosmic reionization) in smaller galaxies.  For galaxies with present 

\vfill

\begin{figure}[h]

\includegraphics{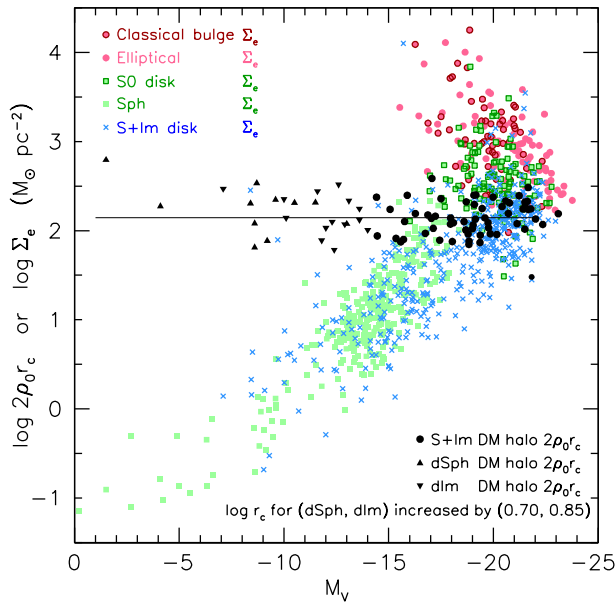}

\caption{\lineskip=0pt \lineskiplimit=0pt
Comparison of DM halo parameters from Fig.~2 with visible matter galaxy parameters 
from Kormendy \& Bender (2012).  Central projected densities are plotted for DM halos; 
effective surface densities $\Sigma_e = \Sigma(r_e)$ are shown for visible components.  
Here $r_e$ is the radius that encloses half of the light of the component.
To convert surface brightnesses to stellar surface densities, we assume mass-to-light ratios 
$M/L_V = 8$ for ellipticals,
         5  for bulges and S0 disks, and
         2  for spiral galaxy disks, Im galaxies, and Sph galaxies.
From Kormendy \& Freeman (2014).
}

\end{figure}

\eject

\noindent $M_V \sim -10 \pm 3$, the baryon depletion is estimated by $\Delta M_B$ in the caption of Figure\ts2.
If the stellar $M/L_V$ is similar in dwarfs and Sc{\ts}--{\ts}Im systems, then $\Delta M_B$ 
measures the mass of stars in dwarfs {\it relative to the mass of stars in Sc--Im galaxies with halos 
of similar~$\rho_\circ$.}  This agrees with their mass-to-light ratios, $M/L_V \sim 10^2$.  
These dwarfs are almost dark, probably because they lost more baryons than the (also substantial) 
loss from Sc\ts--{\ts}Im galaxies.  In dIm galaxies, some baryons are still in cold gas,
but this effect is fairly small.

     Also in Figure 3, projected (not volume!)~DM density $\Sigma_{DM\circ}$ is essentially 
independent of galaxy luminosity $L_V$.  This is a well known result 
(Kormendy \& Freeman 2004;
Spano et al.~2008;
Gentile et al.~2009;
Donato et al.~2009; 
Plana et al.~2010).
It implies a Faber-Jackson (1976) relation of the form DM mass $M_{DM} \propto \sigma^4$.

      Finally, in Figure 3, bulges and elliptical galaxies have $\Sigma_e > \Sigma_{DM\circ}$, 
more so at lower $L_V$.
Kormendy \etal (2009) and Kormendy \& Bender (2012) suggest that they form a sequence of increasing 
dissipation in the formation of smaller galaxies.

      Perhaps the most remarkable result in Kormendy \& Freeman (2014) is shown in Fig.~4.
It shows that rotation-curve decompositions reveal a robust, linear correlation between the
maximum rotation velocity $V_{\rm circ,disk}$ of baryonic disks and the outer circular velocity $V_{\rm circ}$ 
of test particles in their DM halos.  It explicitly shows that $V_{\rm circ,disk} \rightarrow 0$ 
km s$^{-1}$~at $V_{\rm circ}$\ts$>$\ts0{\ts}km{\ts}s$^{-1}$.~In fact, baryons become unimportant at 
$V_{\rm circ} = 42 \pm 4$ km s$^{-1}$.  This $V_{\rm circ,disk} = 0$ intercept agrees very well with the 
typical halo $\sigma$ that we deduced in Section\ts2.  Smaller galaxies are dim or dark.  For example, 
the two extremely faint dSph galaxies in our sample, Segue\ts1 and Coma, have $V_{\rm circ}$ values 
of about 16 km s$^{-1}$.

\vfill

\begin{figure}[h]

\includegraphics{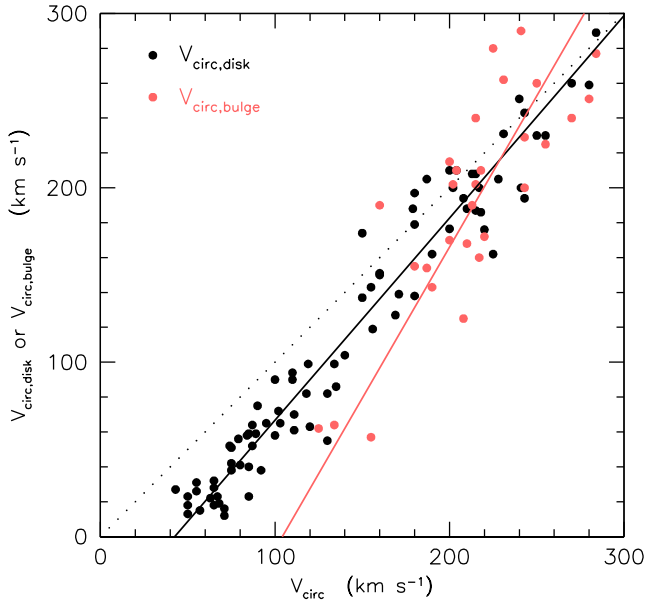}

\caption{\lineskip=0pt \lineskiplimit=0pt
Maximum rotation velocity of the bulge $V_{\rm circ,bulge}$ ({\it red points})
and disk $V_{\rm circ,disk}$ ({\it black points}) given in bulge-disk-halo decompositions of 
galaxy rotation curves whose outer,~DM test particle rotation velocities are $V_{\rm circ}$.  The dotted line
indicates $V_{\rm circ,bulge} = V_{\rm circ,disk} = V_{\rm circ}$.  Every 
red point has a corresponding black point, but many galaxies are bulgeless, and then
only a disk was included in the decomposition.  
This figure shows that the ``rotation~curve~conspiracy'',
$V_{\rm circ,bulge} \simeq V_{\rm circ,disk} \simeq V_{\rm circ}$ for the halo
(Bahcall \& Casertano 1985; 
van Albada \& Sancisi 1986;
Sancisi \& van Albada 1987), 
happens mostly for galaxies with $V_{\rm circ} \sim 200$~km~s$^{-1}$.    The lines are least-squares 
fits with each variable symmetrized around 200 km s$^{-1}$.  The correlation for bulges is steeper
than the one for disks; bulges disappear at $V_{\rm circ} \sim 104 \pm 16$ km s$^{-1}$.  
Disks disappear robustly at $V_{\rm circ} = 42 \pm 4$ km s$^{-1}$.  From Kormendy \& Freeman (2014).  
\lineskip=-20pt \lineskiplimit=-20pt
}

\end{figure}

\eject

\section{Does There Exist a Large Population of Dark Galaxies?}

      From the above results, we conclude that the range of visible matter content of 
$\sigma \sim 30$ km s$^{-1}$ DM halos is large.  The ones with the most baryons rotate enough
to allow rotation curve decomposition.  The ones with the fewest baryons are barely discoverable.
None of these galaxies ``know'' that they must retain $\sim$\ts1\ts\% of their baryons to be
discoverable by us almost 14 billion years after they formed.  Moreover, as luminosity decreases
toward barely discoverable galaxies, these dwarfs become much more numerous as well as much more nearly 
dominated by DM.  And baryon depletion processes should be more efficient in smaller galaxies.  All
this suggests that there may exist a large population of objects that are even darker -- too dark 
to be discovered by current techniques.  This would help to solve the problem that such objects are 
predicted by cold DM theory but not seen in Local-Group-like environments
(Moore \etal 1999;
Klypin \etal 1999).

\section*{Acknowledgments}

      JK thanks the Directors and staff of Mt.~Stromlo Observatory (Australia), of the Observatory of
the Ludwig-Maximilians-University (Munich, Germany), and of the Max-Planck-Institute for Extraterrestrial
Physics (Garching-by-Munich, Germany) for hospitality and support during many visits~over~many
years when this work was done.  We made extensive use of NASA's Astrophysics Data System bibliographic 
services and of the NASA/IPAC Extragalactic Database (NED).  NED is operated by the Jet Propulsion Laboratory 
and Cal Tech under contract with NASA.  And we used the HyperLeda electronic database at 
{\tt http://leda.univ-lyon1.fr}
(Paturel \etal 2003).  JK's work was supported by NSF grants AST-9219221 and AST-0607490, by 
the Alexander von Humboldt-Stiftung (Germany), and by Sonderforschungsbereich 375 of the German Science 
Foundation.  The visits to Mt.~Stromlo were made possible by the long-term support provided 
to JK by the Curtis T.~Vaughan, Jr.~Centennial Chair in Astronomy.

\end{document}